\input{aipcheck}

\documentclass[
    ,final            
  ]
  {aipproc}

\layoutstyle{6x9}

\usepackage{amsmath}
\newcommand{\qbar}{\bar{q}}
\newcommand{\sbar}{\bar{s}}
\newcommand{\cbar}{\bar{c}}

\newcommand{\Ex}[2]{\ifmmode{#1\times10^{#2}}\else{$#1\times10^{#2}$}\fi}

\newcommand{\Psfig}[2]{\includegraphics[width=#1]{#2}}
\newcommand{\Comment}[1]{}

\begin{document}

\title{Exotics from Heavy Ion Collisions}


%
%
%
\author{Akira Ohnishi}{
  address={Yukawa Institute for Theoretical Physics, Kyoto University, Kyoto 606-8502, Japan}
}
\author{Sungtae Cho}{
address={Institute of Physics and Applied Physics, Yonsei University,
Seoul 120-749, Korea}
}
\author{Takenori Furumoto}{
  address={Yukawa Institute for Theoretical Physics, Kyoto University, Kyoto 606-8502, Japan}
 ,altaddress={RIKEN Nishina Center, Hirosawa 2-1, Wako, Saitama 351-0198, Japan}
}
\author{Tetsuo Hyodo}{
  address={Department of Physics, Tokyo Institute of Technology, Meguro 152-8551, Japan}
}
\author{Daisuke Jido}{
  address={Yukawa Institute for Theoretical Physics, Kyoto University, Kyoto 606-8502, Japan}
}
\author{Che Ming Ko}{
  address={Cyclotron Inst. and  Dep. of Physics and Astronomy, Texas A\&M Univ., Texas 77843, U.S.A.}
}
\author{Su Houng~Lee}{
  address={Institute of Physics and Applied Physics, Yonsei University, Seoul 120-749, Korea}
  ,altaddress={Yukawa Institute for Theoretical Physics, Kyoto University, Kyoto 606-8502, Japan}
}
\author{Marina Nielsen}{
  address={Instituto de F\'{\i}sica, Universidade de S\~{a}o Paulo, C.P. 66318, 05389-970 S\~{a}o Paulo, SP, Brazil}
}
\author{Takayasu Sekihara}{
  address={Yukawa Institute for Theoretical Physics, Kyoto University, Kyoto 606-8502, Japan}
  ,altaddress={Department of Physics, Graduate School of Science, Kyoto University, Kyoto 606-8502, Japan}
}
\author{Shigehiro Yasui}{
  address={Institute of Particle and Nuclear Studies, KEK, Ibaraki 305-0801, Japan
\\[1ex](ExHIC Collaboration)}
}
\author{Koichi Yazaki}{
  address={Yukawa Institute for Theoretical Physics, Kyoto University, Kyoto 606-8502, Japan}
  ,altaddress={RIKEN Nishina Center, Hirosawa 2-1, Wako, Saitama 351-0198, Japan}
}

\begin{abstract}
Discriminating hadronic molecular and multi-quark states is a long standing problem in hadronic physics.
We propose here to utilize relativistic heavy ion collisions to resolve this problem, as exotic hadron yields are expected to be strongly affected by their structures.
Using the coalescence model, we find that the exotic hadron yield relative to the statistical model result is typically an order of magnitude smaller for a compact multi-quark state,
and larger by a factor of two or more for a loosely bound hadronic molecule.
We further find that some of the newly proposed heavy exotic states
could be produced and realistically measured at RHIC and LHC.
\end{abstract}

\maketitle



We are now in a new stage of hadron physics,
where various exotic hadron candidates have been discovered
starting from a penta-quark state $\Theta^+(1530)$~\cite{Nakano:2003qx}
and tetra-quark states, $D_{sJ}(2317)$~\cite{Aubert:2003fg}
and $X(3872)$~\cite{Choi:2003ue}.
We cannot properly explain these states within the simple quark model,
then multi-quark and/or hadronic molecule components would be expected.
An important aspect of exotic hadron physics thus involves the discrimination
between a compact multi-quark configuration and a loosely bound molecular
configuration.
We have recently found that hadron yields in relativistic heavy ion collisions
could provide useful information to address this question~\cite{ExHIC}.
The hadron yield in the coalescence model relative to the statistical model
result is found to be smaller in a compact multi-quark configuration
because of the suppressed coalescence probability from additional quarks,
and larger in a loosely bound hadronic molecule state
because of the large size in which constituent hadrons can coalesce.


In the statistical model, the number of produced hadrons
of a given type $h$
per unit rapidity is given by~\cite{Andronic:2005yp}
\begin{align}
\label{Eq:Stat}
N_h^\mathrm{stat} = & V_H \frac{g_h}{2 \pi^2}
\int_0^\infty \frac{p^2 dp}{\gamma_h^{-1}e^{(E_h-B_h\mu_B-S_h\mu_S)/T_H} \pm 1}\ ,
\end{align}
where $g_h$, $\gamma_h$, $B_h$, $S_h$ are, respectively,
the degeneracy, fugacity, baryon number and strangeness of the hadron.
Following the expanding fire-cylinder model~\cite{Chen:2003tn} for central Au+Au collisions at $\sqrt{s_{NN}}=200$ GeV at RHIC,
the volume, temperature, and baryon and strangeness chemical potentials of the source at statistical hadron production
are taken to be
$V_H=1908~\mathrm{fm}^3, T_H=175~\mathrm{MeV}, \mu_B=20~\mathrm{MeV}$
and $\mu_S=10~\mathrm{MeV}$, respectively.
The fugacity $\gamma_c=6.40$ is introduced for $c$ and $\bar{c}$ quarks,
to reproduce the expected total charm quark number $N_c=3$.

In the coalescence model~\cite{Chen:2003tn,Coal},
the hadron yield of type $h$ containing $n$ constituents at mid-rapidity is obtained using harmonic oscillator (Gaussian) wave functions as,
\begin{align}
\label{Eq:Coal}
N_h^{\rm coal} \simeq& g_h\prod_{j=1}^n \frac{N_j}{g_j}
\prod_{i=1}^{n-1}
\frac{(4\pi\sigma_i^2)^{3/2}}{V(1+2\mu_iT\sigma_i^2)} \left[
\frac{4\mu_i T\sigma_i^2}{3(1+2\mu_{i}T\sigma_{i}^{2})}
\right]^{l_i}\; ,
\end{align}
where $N_j$ ($g_j$) is the number (degeneracy) of the $j$-th constituent,
$\mu_i$ ($l_i$) is the reduced mass (orbital angular momentum)
for the $i$-th Jacobi coordinate, and $\sigma_i=1/\sqrt{\mu_i \omega}$
with $\omega$ being the oscillator frequency.
Taking the quark numbers $N_u=N_d=245$ and $N_s=150$
at hadronization ($T=175~\mathrm{MeV}, V=1000~\mathrm{fm}^3$)
for RHIC~\cite{Chen:2003tn},
we find that the addition of a $s(p)$-wave quark leads
to the  coalescence factor of 0.360 (0.093).
Hadrons with more constituents or finite orbital angular momentum~\cite{KanadaEn'yo:2006zk}
are hence suppressed.
By fitting the reference \textit{normal} hadron yields 
(such as $\rho$, $\Lambda(1115)$ or $\Lambda_c(2286)$) in the statistical model,
we fix $\omega=550, 519(385)$ MeV for hadrons composed of light quarks,
and light and strange (charm) quarks, respectively.
Weakly bound hadronic molecules are assumed to be formed at the kinetic freezeout point ($T_{F}=125$ MeV, $V_F=11322~\mathrm{fm}^3$).
For a two-body $s$-wave hadronic molecule state,
$\omega$ is determined from the radius ($\omega= 3/(2\mu_R \langle{r^2}\rangle)$)
or the binding energy ($\omega=6\times\mathrm{B.E.}$).

\begin{table}[htdp]
\caption{
Quantum numbers, configurations, and oscillator frequencies in hadronic molecule configurations
for exotics discussed in this proceedings.
The undetermined quantum numbers, un-established particles, newly predicted particles,
and $p$-wave coalescence configurations
are marked by $^{*)}$, $^{\ddag)}$, $^{\dag)}$, and $(p)$ respectively.
For hadron molecules, $\omega_\mathrm{Mol.}$ is fixed by the binding energy of hadrons (B),
the inter-hadron distances (R),
or the same as that for the subsystem (T).
$m$ and $\omega_\mathrm{Mol.}$ are given in the unit of MeV.
}
\begin{small}
\begin{tabular}{c|c|c|c|c|c|c|c}
\hline
Particle &$m$ & $I$ & $J\pi$ & $2q/3q/6q$ & $4q/5q/8q$ & Mol.  &$\omega_\mathrm{Mol.}$\\
\hline$f_0(980)$  &980&0 &$0+$    & $q\qbar, s\sbar~(p)$  & $q\qbar s\sbar$ & $K\bar{K}$&67.8(B)\\
$a_0(980)$  &980&1  &$0+$   & $q\qbar~(p)$  & $q\qbar s\sbar$ & $K\bar{K}$        	&67.8(B)\\
$D_s(2317)$ &2317&0  &$0+$   & $c\sbar~(p)$  & $q\qbar c\sbar$ & $DK$          		&273(B)\\
$T_{cc}^1$ $^{\dag)}$  &3797&0  &$1+$   &$-$& $qq\cbar\cbar$ & $\bar{D}\bar{D}^*$    	&476(B)\\
$X(3872)$   &3872&0  &$1+$ $^{*)}$   &$-$& $q\qbar c\cbar$ & $\bar{D}D^*$    		&3.6(B)\\
$Z^+(4430)$ $^{\ddag)}$ &4430&1  &$0-$ $^{*)}$   &$-$& $q\qbar c\cbar~(p)$ & $D_1\bar{D}^*$&13.5(B)\\
\hline $\Lambda(1405)$ &1405&0  &$1/2-$ & $qqs~(p)$     & $qqqs\qbar$   & $\bar{K}N$    &20.5(R)-174(B)\\
$\Theta^+(1530)$ $^{\ddag)}$ &1530&0  &$1/2+$ $^{*)}$ &$-$& $qqqq\sbar~(p)$   &$-$	&$-$\\
$\bar{K}KN$ $^{\dag)}$ &1920&1/2&$1/2+$ & $-$     & $qqqs\sbar~(p)$   & $\bar{K}KN$   &42(R)\\
$\bar{D}N$ $^{\dag)}$  &2790&0  &$1/2-$ &$-$& $qqqq\cbar$   & $\bar{D}N$    		&$-$\\
$\Theta_{cs}$ $^{\dag)}$  &2980&1/2&$1/2+$ &$-$& $qqqs\cbar~(p)$   &$-$			&6.48(R)\\
\hline $H$  $^{\dag)}$   &2245&0  &$0+$   & $qqqqss$  &$-$& $\Xi{N}$  			&73.2(B)\\
$\bar{K}NN$ $^{\ddag)}$ &2352&1/2&$0-$ $^{*)}$ & $qqqqqs~(p)$  & $qqqqqqs\qbar$& $\bar{K}NN$ &20.5-174(T)\\
$\Omega\Omega$ $^{\dag)}$  &3228&0  &$0+$   & $ssssss$  &$-$& $\Omega\Omega$    	&98.8(R)\\
$H_c^{++}$ $^{\dag)}$  &3377&1  &$0+$   & $qqqqsc$  &$-$& $\Xi_cN$    			&187(B)\\
$\bar{D}NN$ $^{\dag)}$ &3734&1/2&$0-$   &$-$& $qqqqqq\,q\cbar$  & $\bar{D}NN$   	&6.48(T)\\
\hline
\end{tabular}
\end{small}
\label{summary}
\end{table}

\begin{figure}
\Psfig{14cm}{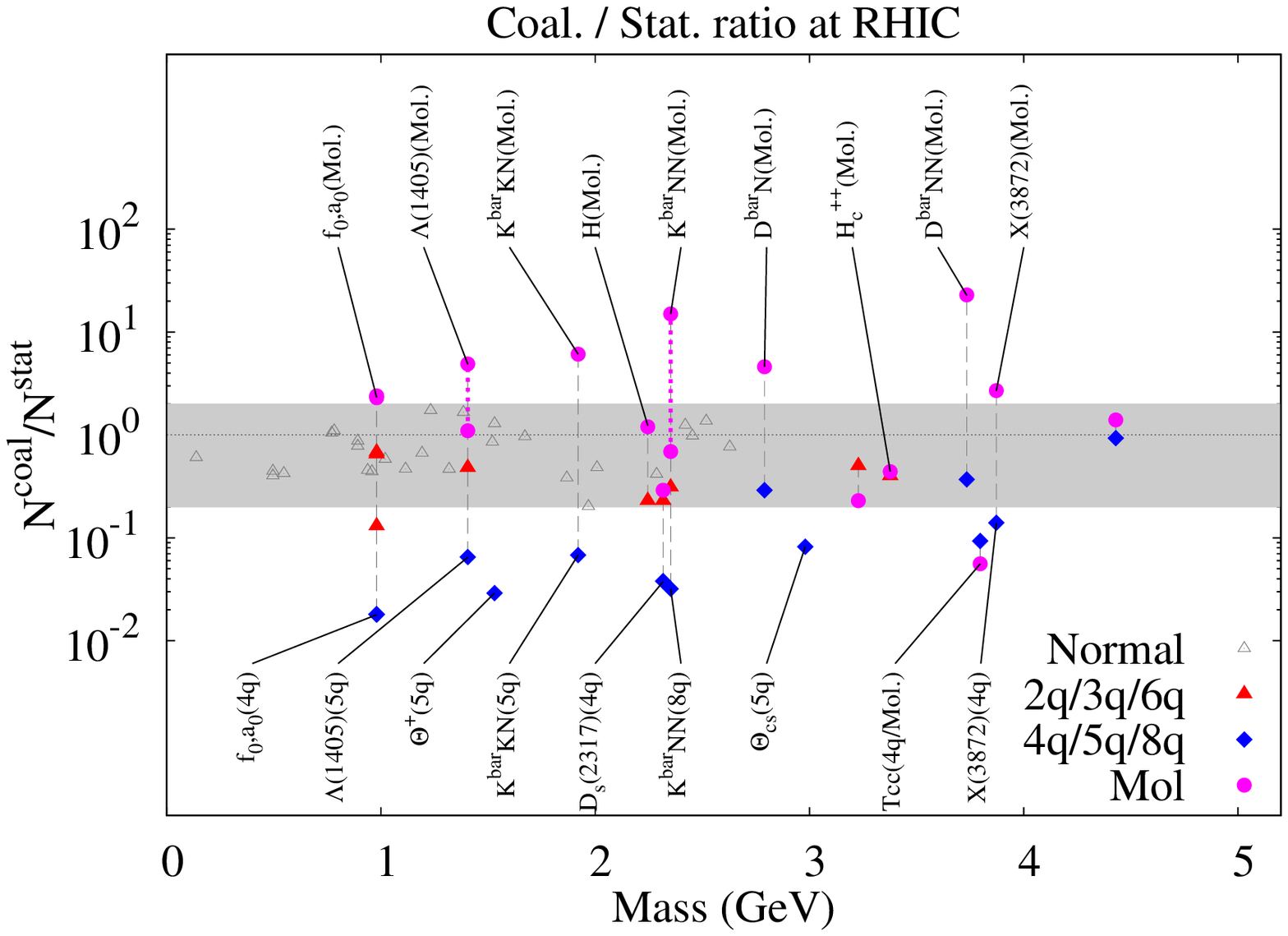}
\caption{(Color online) Multi-quark hadron production at RHIC in the coalescence model relative to the statistical model.
The patterns also holds for LHC as freezeout conditions are similar to that of RHIC.}
\label{Fig:Mass}
\end{figure}

We show the list of hadrons considered here in Table \ref{summary},
including the proposed states,
$T_{cc}^1$~\cite{Zouzou86,Lee09},
$Z^+(4430)$~\cite{BelleZ1},
$\bar{K}KN$\cite{Jido-Enyo},
$\bar{D}N$ and $\bar{D}NN$~\cite{Yasui:2009bz},
$\Theta_{cs}$~\cite{Lipkin},
$H$~\cite{Jaffe76},
$\bar{K}NN$\cite{Akaishi},
$\Omega\Omega$~\cite{di-omega},
and
$H_c^{++}$~\cite{Lee09}.
In Fig.~\ref{Fig:Mass},
we show the ratio $R_h=N_h^\mathrm{coal}/N_h^\mathrm{stat}$
between the yields in the coalescence and statistical models.
%
The grey band ($0.2 < R_h < 2$) covers the normal hadron ratios with $2q$ and $3q$ (open triangles),
including the crypto-exotic hadrons with usual $2q/3q$ configurations.
The ratio is below the normal band ($R_h < 0.2$) for a compact multi-quark configuration (diamonds).
In particular, for light multi-quark configurations such as 
the tetraquark configurations of $f_0(980)$ and $a_0(980)$,
the ratios are an order of magnitude smaller than those of normal hadrons or molecular configurations.
This is consistent with a naive expectation that the probability to combine $n$-quarks into a compact region is suppressed as $n$ increases.
This suppression also applies to $5q$ states in multi-quark hadrons
($\Lambda(1405)$, $\Theta^+(1530)$ and $\bar{K}KN)$ and the $8q$ state in $\bar{K}NN$.
The ratios obtained by assuming hadronic molecular configurations (circles)
are found to lie mostly above the normal band ($R_h > 2$).
Moreover, these ratios depend on the size of the hadronic molecule; loosely bound extended molecules with larger size would be formed more abundantly.
For $\Lambda(1405)$ as an example,
we find $R_h=1.1$ for a small size ($\omega=174$ MeV)
obtained from the relation between the binding energy and $\omega$,
while a coupled channel analysis~\cite{Lambda1405} gives a larger $\langle r^2 \rangle$ and hence a small $\omega$ ($=20.5$ MeV),
leading thus to a larger $R_h=4.9$.

We conclude from the above discussions that the yield of a hadron in relativistic heavy ion collisions reflects its structure and thus can be used as a new method to discriminate the different pictures for the structures of multi-quark hadrons.
Indeed, based on the ratio $f_0(980)/\rho^0 \sim 0.2$ from the preliminary measurement of the STAR Collaboration at RHIC~\cite{STAR_f0}
and the statistical model prediction $\rho^0=42$,
we find the number $f_0(980) \sim 8$.
This is enhanced from the statistical model result ($\sim 5.6$).
Comparing with the coalescence model results
(3.8, 0.73, 0.10, 13 for $q\qbar, s\sbar, q\qbar{s}\sbar, K\bar{K}$),
the measured yield is consistent with the picture that the $f_0(980)$ has
substantial $K\bar{K}$ components,
and a pure tetraquark configuration can be ruled out for its structure.
Further experimental effort to reduce the error is therefore highly desirable
in understanding the structure of $f_0(980)$ and to put an end to this highly
controversial issue.
Similarly, efforts to measure the yields of other hadrons
and newly proposed exotic states listed in Table \ref{summary} will provide new insights to a long standing challenge in hadron physics.


\begin{theacknowledgments}
This work was initiated at the workshop on 
"Exotics from Heavy Ion Collisions",
at the Yukawa Institute of Theoretical Physics.
This work was supported in part by
the Korean BK21 Program and KRF-2006-C00011,
KAKENHI (Nos. 21840026, 22105507 and 22-3389, 21105006 and 22105514),
the global COE programs (Kyoto U. and Tokyo Inst. of Tech.) from MEXT,
the U.S. National Science Foundation under Grant No. PHY-0758115,
the Welch Foundation under Grant No. A-1358, and the CNPq and FAPESP.

\end{theacknowledgments}



\bibliographystyle{aipproc}   


\end{document}